\documentclass[journal]{IEEEtran}

\IEEEoverridecommandlockouts
\usepackage{cite}
\usepackage{amsmath,amssymb,amsfonts}
\usepackage{algorithm}
\usepackage{algorithmic}

\usepackage{graphicx}
\usepackage{amsfonts}
\usepackage{booktabs}
\usepackage{lipsum}
\usepackage[table,xcdraw]{xcolor}
\usepackage{xcolor}
\usepackage[letterpaper, left=0.75in, right=0.75in, bottom=1.05in,
top=1in]{geometry}

\usepackage{graphicx}
	\ifCLASSOPTIONcompsoc
    	\usepackage[caption=false, font=normalsize, labelfont=sf, textfont=sf]	{subfig}
	\else
\usepackage[caption=false, font=footnotesize]{subfig}
	\fi
	
\usepackage[marginal]{footmisc}
\usepackage{textcomp}
\usepackage{multirow}

\usepackage{comment}
\usepackage{blindtext}
\def\BibTeX{{\rm B\kersn-.05em{\sc i\kern-.025em b}\kern-.08em
  T\kern-.1667em\lower.7ex\hbox{E}\kern-.125emX}}

\ifCLASSINFOpdf
\else
\fi
 
\begin{document}
\title{A C-V2X Platform Using Transportation Data and Spectrum-Aware Sidelink Access}

\author{
\IEEEauthorblockN{\textbf{Chia-Hung Lin}\IEEEauthorrefmark{1}, \textbf{Shih-Chun Lin}\IEEEauthorrefmark{1}, \textbf{Chien-Yuan Wang}\IEEEauthorrefmark{1} and \textbf{Thomas Chase}}\IEEEauthorrefmark{2}\\
\vspace{1em}
\IEEEauthorblockA{\IEEEauthorrefmark{1}Intelligent Wireless Networking Laboratory, Department of Electrical and Computer Engineering,\\ 
North Carolina State University, Raleigh, NC 27695\\}
\IEEEauthorblockA{\IEEEauthorrefmark{2}Institute for Transportation Research and Education, North Carolina State University, Raleigh, NC 27695\\}
{Email: clin25@ncsu.edu; slin23@ncsu.edu; cwang64@ncsu.edu; rtchase@ncsu.edu} \vspace{-0.35 in}
}
\maketitle

\begin{abstract}
Intelligent transportation systems and autonomous vehicles are expected to bring new experiences with enhanced efficiency and safety to road users in the near future. However, an efficient and robust vehicular communication system should act as a strong backbone to offer the needed infrastructure connectivity. Deep learning (DL)-based algorithms are widely adopted recently in various vehicular communication applications due to their achieved low latency and fast reconfiguration properties. Yet, collecting actual and sufficient transportation data to train DL-based vehicular communication models is costly and complex. This paper introduces a cellular vehicle-to-everything (C-V2X) verification platform based on an actual traffic simulator and spectrum-aware access. This integrated platform can generate realistic transportation and communication data, benefiting the development and adaptivity of DL-based solutions. Accordingly, vehicular spectrum recognition and management are further investigated to demonstrate the potentials of dynamic slidelink access. Numerical results show that our platform can effectively train and realize DL-based C-V2X algorithms. The developed slidelink communication can adopt different operating bands with remarkable spectrum detection performance, validating its practicality in real-world vehicular environments.
\vspace{1.5ex}
\end{abstract}

\IEEEpeerreviewmaketitle

\section{Introduction}
With the promising connectivity provided by next-generation communication systems, such as 5G and 6G, several novel applications are developing in full flourish. Among those applications, intelligent transportation system (ITS) and autonomous vehicles are anticipated to bring new experiences with enhanced efficiency and safety to road users in the near future \cite{ITSsurvey2}. In macroscopic scale, with the support of ITS algorithms, the data collected from road users can be used to perform traffic flow prediction, optimal path planning, traffic light control, and so on, leading to the improved transportation efficiency. In microscopic scale, via exchanging information with surrounding vehicles and infrastructures, autonomous vehicles can be employed to assist drivers to mitigate fatigue and increase safety during driving. However, an efficient and robust vehicular communication system should act as a strong backbone to offer the needed infrastructure connectivity of ITS and autonomous vehicles \cite{VCsurvey}.
In light of this situation, vehicular communications is a important but challenging research topic to communication society, bringing several special challenges to transitional communication algorithms \cite{VCsurvey}. Specifically, first, ultra-reliable low latency communications (URLLC) is an important consideration to vehicular communications due to its short reaction-time nature. 
Secondly, as the high mobility feature of vehicular communications, algorithms should be able to adopt different environments via fast reconfiguration to better serve vehicular communications. 
As a result, developing suitable algorithms for vehicular communications, which satisfies the aforementioned requirements, is not trivial and challenging. 

\begin{figure*}
    \centering
    \includegraphics[width=0.9\linewidth]{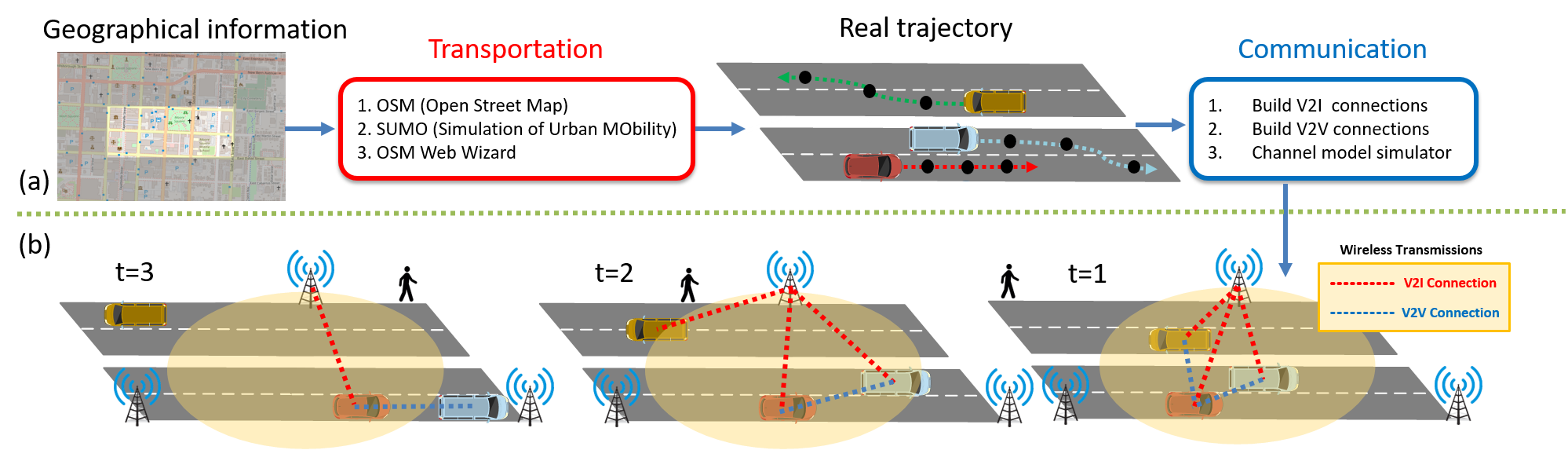}
    \caption{The Illustration of the proposed vehicular communication platform. In the proposed platform, transportation module and communication module are integrated to generate realistic data for verification of vehicular communication algorithms. To be more specific, when a geographical information is imported, transportation module will create realistic traffic on it, then wireless communications will be simulated. As a result, the proposed platform can be used to assess the achieved performance of various vehicular communication algorithms in real scenarios.}
    \label{fig:Reconstruction}
\end{figure*}

In the past few years, researchers focus on the development of deep learning (DL)-based solutions for vehicular communication. Compared to traditional optimization-based algorithms, which often involves complex matrix operations and iterations, DL models can be used to construct efficient algorithm with URLLC and adaptability since exhausted computation tasks can be finished in offline training phase. Hence, DL-based algorithms are widely adopted in various vehicular communications applications \cite{ITSsurvey2}. 
However, to train DL-based algorithms, abundant amount of data with high quality should be provided. When it comes to the vehicular communications, collecting sufficient amount of real transportation data is costly and difficult. Many works are forced to use simulation data to finish experiments, limiting the power of DL-based algorithms due to the absence of critical features in practical and also raising doubts about its practicality \cite{GYL}. 
To tackle this dilemma, in this paper, we propose a vehicular communication platform based on real traffic simulator. As shown in Fig.~\ref{fig:Reconstruction}, the proposed platform integrates both transportation and communication modules for the verification of various wireless communication and networking algorithms.
Specifically, several open source software, including Simulation of Urban MObility (SUMO), openStreetMap (OSM), and OSM Web Wizard \cite{SUMO}, are combined in the transportation module. By doing so, in our transportation module, customized geographic information can be imported as a region of interest first, then realistic trajectories of each vehicle will be created on the region of interest based on adjustable parameters. Given those realistic trajectories, in our communication module, vehicle-to-vehicle (V2V) and vehicle-to-infrastructure (V2I) connections, which is simulated based on latest specifications and papers, are created occasionally to simulate vehicular communications environment. Moreover, we offer sub6GHz and Terahertz (THz) wireless communications for users to choose from now. 
Different from existing simulation platforms, the trajectories generated from our transportation module will be affected by traffic light and the actions of surrounding vehicles, offering realistic and critical patterns. At the same time, given a interested scenario, the proposed platform can be used to generate huge amount of data quickly, stimulating the development of DL-based vehicular communications algorithms.  
To validate the proposed platform, we develop a DL-based spectrum management algorithm for sub6GHz vehicular communications. Moreover, the ability of the proposed algorithm to adopt different communication scenarios is also investigated in detail. To explain, spectrum management is an indispensable component to realize efficient vehicular communications. By employing spectrum sensing algorithms in vehicular communications, underutilized spectrum can be reused and consequently leads to a better overall spectrum efficiency \cite{Lin3}. Also, an effective spectrum sensing algorithm can be used as a preceding algorithm to perform critical radio resources allocations in vehicular communications. When it comes to heterogeneous communication systems or military usage \cite{SS4}, spectrum management also shows its importance in terms of provided efficiency improvement. Therefore, we implement critical spectrum sensing algorithms in our platform for further verification.
We summarize the major contributions of this work as follows:
\begin{itemize}
  \item A vehicular communication verification platform is proposed to demonstrate vehicular communication algorithms and assess their performance in real scenarios. To the best of our knowledge, this work is the first to provide an integrated vehicular platform with real traffic simulator and next generation communication systems following specifications and standards. Specifically, realistic traffic in customized region of interest can be created by the transportation module. Then communication connections will be conducted via either sub6GHz or THz bands by the communication module. The proposed platform can be used to demonstrate existing vehicular communication algorithms, especially DL-based algorithm to satisfy the need of hung amount of data with high quality.
  \item To demonstrate the developed platform, we develop DL-based spectrum management sub6GHz solution based on our previous work~\cite{Ourwork1}. Note all the communication system settings are following latest standard and specifications to better serve current vehicular communications. The comparison of the proposed DL-based and existing generative adversarial network (GAN)-based algorithms is also conducted in different bands. Simulation results reveal that our spectrum management algorithm can offer excellent performance regardless the operation band, showing the practicality of our spectrum management algorithm. 
\end{itemize}

The rest of paper is organized as follows. Section II introduces the considered spectrum management problem. Section III provides the detail of the developed algorithm. In section III, we first introduce the comprehensive information about the developed platform, and then use the generated data to present and discuss the achieved performance of spectrum management algorithms. Finally, section V concludes the whole paper.

\section{System Model and Problem Description}

\subsection{Signal and Communication Setups} \label{sssec:num1}

Considering a vehicular communications scenario, followed by cellular-V2X (C-V2X) structure with operation frequency band $[W_1,W_2]$ (Hz), a base station (BS) serves several user equipments (UEs) in its cell coverage. To increase spectrum efficiency by performing spectrum management \cite{5GSS}, each transmission time-slot will be divided into two phases: sensing phase and data transmission phase. In sensing phases, an UE, which wishes to create new connection using spectrum holes in the radio environment, will perform spectrum sensing to detect existing connections and seek for idle bands first, then the desired connection can be built in the following data transmission phase. Suppose there are $I$ connections existing in the time-slot, the received signal $\textbf{x}_{c}(t)$ at the UE side can be modeled as 
\begin{equation}
\begin{aligned}
		\textbf{r}_{c}(t) = \textbf{x}_{c}(t) + \textbf{w}(t)    
		= \sum^{I}_{n=1}\textbf{h}_{n}(t)*\textbf{s}_{n}(t)+\textbf{w}(t),
\end{aligned}
\end{equation}
where $\textbf{s}_{n}(t)$ and $\textbf{h}_{n}(t)$ are the transmitted signal and corresponding channel effect of existing transmissions and $\textbf{w}(t)$ is the noise in receiver side. 
In a given sensing phase, by using an ADC with Nyquist sampling rate $f_{s}$, a perfect discrete time sequence $\textbf{r}[n] = \textbf{x}[n] + \textbf{w}[n] = \textbf{x}_{c}(\frac{n}{f_{s}})+\textbf{w}(\frac{n}{f_{s}}), n = 0,1,...,N_s-1$ can be obtained as $\textbf{r} \in C^{N_s \times 1}$. Also, the corresponding discrete Fourier transformation (DFT) spectrum can be obtained by performing $\textbf{R} = \textbf{F}\textbf{r}$, where $\textbf{F}$ is the DFT basis matrix. However, getting DFT spectrum in this way is costly and impractical. To tackle this drawback, spectrum reconstruction problem will be introduced and formulated.




\subsection{Problem Formulation}

Sub-Nyquist sampling and spectrum reconstruction based on undersampled measurements can be introduced to avoid the exhausted sampling process. 
To explain mathematically, a sub-Nyquist sampling process can be expressed as 
\begin{equation}
\begin{aligned}
		\textbf{y} = \Phi \textbf{Fr},
\end{aligned}
\end{equation}
where $\Phi_{M \times N_s}$ is the sensing matrix, which satisfies $M<N_s$ to perform compression.
Our goal in this paper is to design a way to reconstruct the the desired signal $\textbf{x} \in C^{N_s \times 1}$ or corresponding clean DFT spectrum $\textbf{X} = \textbf{F}\textbf{x}$ from the noisy and undersampled measurements $\textbf{y} \in C^{M \times 1}$.
Traditional compressed sensing algorithms can be used to perform reconstruction by solving the optimization problem:
\begin{equation}
  \begin{aligned}
    & \underset{x}{\text{minimize}}
    & & ||\textbf{x}||_{1} \\
    & \text{subject to}
    & & ||\textbf{y}-\Phi \textbf{Fx}||_{2} \leq \epsilon,
  \end{aligned}
\end{equation}
where $\epsilon$ is the distortion threshold. However, the computational efficiency and execution time can be further improved.
To explain, existing spectrum reconstruction algorithms only focus on the design of reconstruction process and ignore the design of compression process, limiting the achieved efficiency.
As an alternative, we aim to design a joint compression and reconstruction algorithm to improve the overall efficiency.
Hence, instead of solving Eq. (3), the interested joint optimization problem can be expressed as:
\begin{equation}
  \begin{aligned}
    & \underset{f, \Phi}{\text{minimize}}
    & & ||\textbf{X}-f(\Phi \textbf{Fr})||_2 ,
  \end{aligned}
\end{equation}
where $f$ is the reconstruction function. 

\section{The Development of DL-based Spectrum Reconstruction}
To solve the optimization problem in Eq. (4), we develop a DL-based algorithm to perform efficient compression and reconstruction. To explain, designing and evaluating a sensing matrix in the compression process via optimization-based algorithms is not travail, turning out that existing algorithms only focus on solving Eq. (3) instead of solving Eq. (4). In the proposed algorithm, benefiting from the end-to-end learning nature of DL-based algorithm, backprorogation algorithm is used to improve the compression and reconstruction process simultaneously via gradient decent mechanism in each epoch, achieving effective joint optimization. 
As a result of the joint optimization, undersampled measurements with more critical information compared to existing algorithms can be obtained, leading to better efficiency and consequent better reconstruction results. 


\begin{figure*}
    \centering
    \includegraphics[width=0.90\linewidth]{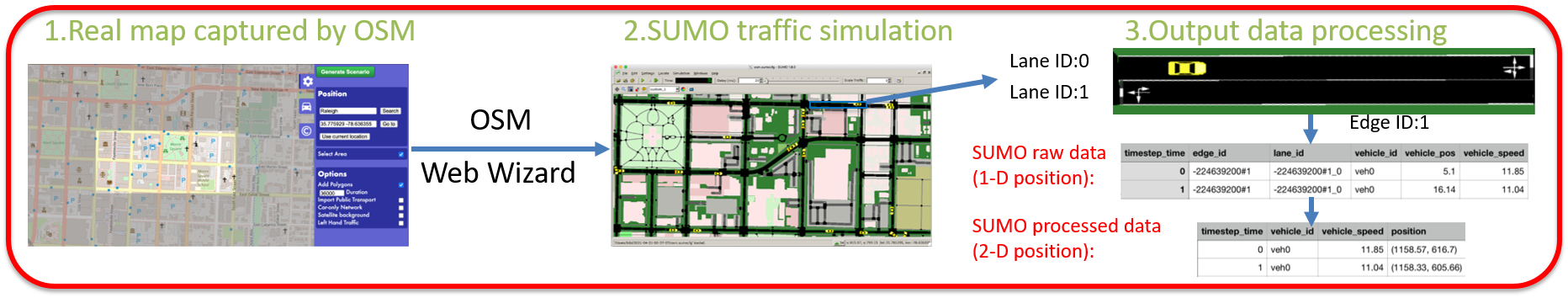}
    \caption{The detailed process of the transportation module: Using the transportation module, realistic traffic will be created in region of interest and the trajectories of vehicles will be recorded for further use.}
    \label{fig:TModule}
\end{figure*}

\subsection{Deep Learning Model Architecture}
We create three consecutive modules (namely compression, coarse reconstruction, and fine reconstruction modules) in our DL model for efficient spectrum reconstruction.
The compression module is a specially-designed one-layer convolutional neural network (CNN) with $M$ filters to produce undersampled measurements by making the trainable weights in the compression module act as the content of sensing matrix, being expressed as:
\begin{equation}
\begin{aligned}
		\textbf{z}_{DL} = \Phi_{DL} \textbf{Fr},
\end{aligned}
\end{equation}
where $\textbf{Fr} \in \mathbb{C}^{N_s}$ is the original spectrum, $\Phi_{DL} \in \mathbb{C}^{M \times N_s}$ is the sensing matrix designed by the compression module, and $\textbf{z}_{DL} \in \mathbb{C}^{M}$ is the obtained undersampled measurements.
To be more specific, the input of the compression module $\textbf{Fr}$ is presented as a real vector with the size of $N_s \times 2$. After the operation of the \textit{compression module}, the output $\textbf{z}_{DL}$ is a real vector with the size of $M \times 2$, standing for the real part and the imaginary part of the undersampled measurements.
Note that there is no activation function in this CNN layer to ensure the whole compression module as a linear operation. After training, the trainable weights in each filter can be represented as a pseudo-random (PN) sequence \cite{DLSS1}. Then received signals will be mixed with $M$ PN sequences and pass through a low-pass filter to get the undersampled measurements $\textbf{z}_{DL}$ in real scenarios.
Once we get the undersampled measurements $\textbf{z}_{DL}$, coarse reconstruction and fine reconstruction module will be employed to perform spectrum reconstruction. The coarse reconstruction is another CNN layer, which has $N_s$ filters with the size of ${M} \times 2$, to generate initial reconstruction. Different from the compression module, batch normalization (BN) and PRelu are employed in this CNN layer to offer nonlinearty. Then the fine reconstruction module, which is developed based on ResNet-structure and has six residual blocks, will be employed to finish final reconstruction. In each residual block, three CNN layers with number of filters 64, 32, and 2, respectively, are built to refine the initial spectrum reconstruction. Behind each layer, we also employ BN and PRelu in the initial reconstruction module to complete the fine-scale reconstruction.

\subsection{Loss Functions}
To leverage the end-to-end training mechanism for solving interested optimization Eq. (4) efficiently, Eq. (4) is employed as as the loss function of the aforementioned model directly. Let $\Theta_{CR}$ stands for the trainable weight in the coarse reconstruction module and $\Theta_{FR}$ represent the trainable weight in the fine reconstruction module and $f(x;\Theta_{CR}. \Theta_{FR})$ is the nonlinear transformation with $\Theta_{CR}$ and $\Theta_{FR}$, the loss function can be expressed as 
\begin{equation}
\begin{aligned}
		Loss = ||\textbf{Fs}-f(\Phi_{DL} \textbf{Fr};\Theta_{CR}. \Theta_{FR})||^2.
\end{aligned}
\end{equation}
During each epoch, $\Phi_{DL}$, $\Theta_{CR}$, and $\Theta_{FR}$ will be updated jointly to minimize the training loss until convergence, generating optimal sensing matrix and trainable weights in each module.
As for the training specifics, the Adam optimizer with initial learning rate $0.0005$ is employed to minimize the loss function. The number of epochs is set as 20 and the mini-batch mechanism is employed with batch size as $128$ to facilitate fast convergence.


\section{A New C-V2X Platform and Numerical Results}
\subsection{Vehicular Communication Platform Construction}
To evaluate the performance of vehicular algorithms in real scenarios, as shown in Fig.~\ref{fig:Reconstruction}, we create a platform integrating real transportation module and spec-defined communication module to obtain simulation results.
\subsubsection{Transportation module}
In order to obtain realistic traffic data, we introduce SUMO~\cite{SUMO} as the transportation module of the proposed platform. To explain, SUMO is an open source platform that can create microscopic realistic traffic data. Moreover, with the OSM combination, we can import the geographic information from real map to evaluate the achieved performance of vehicular communication algorithms in any region of interest of real locations.
As shown in Fig.~\ref{fig:TModule}, given a fixed location, in our transportation module, several parameters can be adjusted to simulate the transportation behavior during different time or events. In this paper, a two direction street with four lanes in Raleigh Downtown, North Carolina is selected as region of interest to investigate the behavior of vehicular communication algorithms in Urban scenario. The area we selected for simulation is around 600k $m^2$ and traffic lights are built and used to control the traffic in this area. The time step is set for 1 second and the average seconds of an new vehicle generated on the map is set as 2.5. Furthermore, we adopt vehicles with 5 $m$ length and 1.8 $m$ width to finish the simulations. The maximum speed of each vehicle is set as 55.56 $m$ per second, and minimum gap of two vehicles is 2.5 $m$. It is noteworthy that the aforementioned parameters can be adjusted flexibly to reflect the traffic in peak hours or non-peak hours or to match any communication specifications or standards.
Moreover, different type of vehicles with different size can be simulated by SUMO platform to generate realistic traffic behavior.
Based on the aforementioned settings, SUMO raw dump traffic data will be created. Then the original one-dimensional position data will be converted to two-dimensional position data first and be passed to communication module for further processing.

\begin{figure*}
    \centering
    \includegraphics[width=0.90\linewidth]{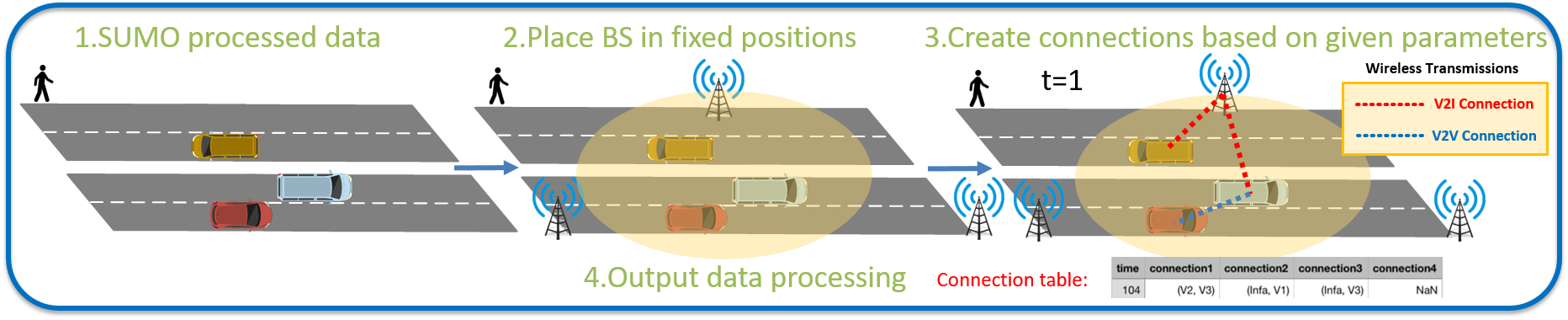}
    \caption{The detailed process of the communication module: In each time step, based on the realistic trajectories data and communication system settings, wireless connections will be created and the detailed information is also provided in our platform.}
    \label{fig:CModule}
\end{figure*}


\subsubsection{Communication module}
After we get the discrete traffic data generated from the transportation module, we follow the latest settings of standards and specifications to construct the communication module. To be more specific, we set one BS on middle of the road. The distance between the V2I connections (i.e., BS and any vehicles) and V2V connections (i.e., any pairs with two vehicles) can be calculated in each time step. In the coverage of the BS or the maximum communication range of V2V connections, we set that for each vehicle there is 0.9 chance that it is communicating with the BS and for each pair of vehicles there is 0.5 chance that they are communicating with each other to create V2V and V2I connections.
Next, it is noteworthy that we offer different operation band options with realistic channel behavior to perform simulations in our communication module. Specifically, in Release 17 \cite{spec1} announced in December 2019, frequency range 1 (sub6GHz) is still considered to support C-V2X communications. We further consider THz band as an option in our platform for next-generation communications. The comparison of achieved performance in those bands can act as a useful reference for system designer to determine the actually band for different applications. 
We set $W_1$ and $W_2$ as $[0,2GHz]$ according \cite{WSS1} to simulate wideband scenario in sub6GHz band.
When it comes to the THz band, we follow our previous paper \cite{Ourwork1} to set $W_1$ and $W_2$ as $[0.1THz, 0.55THz]$.
As for the coverage of V2V and V2I connections, we set that as 100m for sub6GHz band and 15m for THz band based on latest papers.
In summary, there are 2GHz and 450GHz available bandwidth for the usages of sub6GHz and THz transmission, respectively. 

As for the channel effect, we consider frequency and distance-dependent path loss in both bands following the suggestions of 3GPP simulation guideline and latest papers \cite{spec1}.
Specifically, the block slow fading channel with the path loss: $127+2*log_{10}(f)+30*log_{10}(d)$, where $d$ (km) is the distance and $f$ is the center frequency of each connections, is set for sub6GHz band. As for THz band, we follow the setting in our previous paper \cite{Ourwork1} to simulate realistic THz channel behavior with the consideration of non-ideal conditions, such as temperature, atmospheric pressure, air density and non-line of sight effect.
The whole simulation process of communication module is shown in Fig.~\ref{fig:CModule}, in each time step, different number of connections will be created based on the given trajectories and aforementioned communication parameters. Moreover, the detailed information about the transmitter and receiver of each connection is also provided in the proposed platform.


\subsubsection{Data set generation}
By integrating the above two modules, we generate realistic data set for the training of DL-based algorithms. 
Specifically, different from existing works, which employs simple transportation model to describe the vehicle mobility, the trajectories of each vehicle are affected by adjacent vehicles and traffic lights on road in our simulations, leading to more realistic simulation results. 
Based on these trajectories, V2V and V2I connections are built occasionally in our communication module. As compared to latest spectrum reconstruction algorithms, which assumes fixed number of existing connections in their simulations, it is noteworthy that the number of valid connections may change in each sample in our simulations, challenging the ability of DL-based reconstruction algorithms but being more similar to real scenarios. 
When all the communication connections are generated, we assume $N_s = 256$ subcarriers equally spaced within the available bandwidth. We further assume that each existing users chose a random, non-overlapping group of 5 subcarriers to transmit on, with at least 1 subcarriers of guard on either side. Finally, antenna gains from transmitter $G_t$ and receiver $G_r$ are set as 0 dbi for sub6GHz and 50 dbi for THz bands, respectively for fair comparison. 
Although we consider the single input single output structure for the communication system design of each band, the achieved results should be the same for multiple input multiple output structure as the only difference is the provided array gain in our interested problem. 
For each operating band, we generate 50000, 10000, and 10000 samples and corresponding labels for training, validation, and testing sets, respectively. Note that all the inputs and labels are normalized as [1,0] to prevent computational issues of DL-based algorithms. All the reported results are the average values over the testing set.



\subsection{Performance Metrics}
We assess different algorithms from both machine learning and communication perspectives in this paper.
As for the machine learning performance metrics, mean-square-error (MSE), cosine similarity, and structure similarity (SSIM) between original spectrum and reconstructed results, are reported in this paper \cite{Ourwork1}. 
As for the communication performance metrics, we combine the output of different spectrum reconstruction algorithms with simple energy detection mechanism to perform spectrum sensing. In both original and reconstruction spectrum, a subcarrier will be considered as occupied if the power of it exceeds 0.5. Then detection rate $P_d$ and false alarm rate $P_f$ can be calculated to reflect the spectrum sensing performance, Note that the better trade-off between $P_d$ and $P_f$ implies the better spectrum efficiency and better communication quality as new users can always find spectrum holes to conduct interference-free data transmission.

\subsection{Performance Comparison}
Table I reflects the achieved performance of different algorithms in different operating bands with signal-to-noise ratio = 30 dB and compression rate = 0.125. In both sub6GHz and THz bands, the proposed algorithm outperforms GAN-based algorithm in various performance metrics significantly. To be more specific, in terms of machine learning performance metrics, the proposed algorithm can offer excellent reconstruction results, which is very close to the original spectrum. Moreover, the proposed algorithm can strike a good trade-off between detection rate and false alarm rate to aid vehicular communications. On the other hand, the reconstruction quality provided by GAN-based algorithm is unacceptable, failing to reflect the utilizing situations of original spectrum and failing to perform effective spectrum management. Specifically, it is noteworthy that the detection rate of GAN-based algorithm is far less than $90\%$, which is unsatisfactory for any communication systems.  
It is also noteworthy that the simulation results reveal that the proposed spectrum reconstruction can be employed in different bands as its data-driven nature can adopt different channel characteristics automatically. 
Finally, although THz bands can provide abundant bandwidth to support data rate-demanding or low latency applications, we report that the extra power consumption or advanced technologies (i.e., beamforming technique, antenna design) should be used to overcome the higher path loss compared to sub6GHz band.

\begin{table}[]
\caption {Comparison of spectrum reconstruction in different bands with SNR = 30 dB and compression rate = 0.125}

\centering
\begin{tabular}{ccccc}
\hline
\multicolumn{1}{c|}{}                                                                                                                                    & \multicolumn{2}{c|}{GAN}                                      & \multicolumn{2}{c}{Proposed}                                                               \\ \hline
\multicolumn{1}{c|}{}                                                                                                                                    & \multicolumn{1}{c|}{sub6GHz}  & \multicolumn{1}{c|}{THz}      & \multicolumn{1}{c|}{sub6GHz}                           & THz                                \\ \hline
\multicolumn{5}{c}{\textbf{Machine learning performance metrics}}                                                                                                                                                                                                                                                      \\ \hline
\multicolumn{1}{c|}{MSE}                                                                                                                                 & \multicolumn{1}{c|}{0.0685}   & \multicolumn{1}{c|}{0.0187}   & \multicolumn{1}{c|}{\textbf{0.0026}}  & \textbf{9.61e-04} \\ \hline
\multicolumn{1}{c|}{\begin{tabular}[c]{@{}c@{}}Cosine\\ Similarity\end{tabular}} & \multicolumn{1}{c|}{0.3370}   & \multicolumn{1}{c|}{0.4394}   & \multicolumn{1}{c|}{\textbf{0.9951}}  & \textbf{0.9908}   \\ \hline
\multicolumn{1}{c|}{SSIM}                                                                                                                                & \multicolumn{1}{c|}{0.3033}   & \multicolumn{1}{c|}{0.6289}   & \multicolumn{1}{c|}{\textbf{0.8523}}  & \textbf{0.9386}   \\ \hline
\multicolumn{5}{c}{\textbf{Communication performance metrics}}                                                                                                                                                                                                                                                         \\ \hline
\multicolumn{1}{c|}{$P_d$}                                                                                                                               & \multicolumn{1}{c|}{2.47e-04} & \multicolumn{1}{c|}{0.0622}   & \multicolumn{1}{c|}{\textbf{0.9}}     & \textbf{0.9449}   \\ \hline
\multicolumn{1}{c|}{$P_f$}                                                                                                                               & \multicolumn{1}{c|}{0}        & \multicolumn{1}{c|}{2.68e-04} & \multicolumn{1}{c|}{\textbf{2.6e-04}} & \textbf{0.0014}   \\ \hline
\end{tabular}
\end{table}

\section{Conclusion}
We propose a vehicular communication verification platform to facilitate the development of DL-based C-V2X communication. A transportation module based on SUMO for realistic transportation data and a communication module based on the latest specifications are integrated to offer practical simulation results.
Using the proposed platform, we further develop a DL-based spectrum management algorithm for the sub6GHz band and investigate the adaptability with actual transportation data and different communication settings. Simulation results confirm that the proposed algorithm can be employed in various communication systems to offer impressive results. 

For future work, we plan to add other 3GPP communication options, such as millimeter wave, and connection types, such as vehicle-to-pedestrian and vehicle-to-network connections. Moreover, we will create computation module to address computation-related issues, including edge computing, on-device learning, and distributed machine learning training in vehicular communications.  



\ifCLASSOPTIONcaptionsoff
  \newpage
\fi

\bibliography{references}
\bibliographystyle{IEEEtran}

\end{document}